\documentclass[a4paper,11pt]{article}
\usepackage{Layout_stuff/pos}
\usepackage{bbold}
\usepackage{amsmath}
\usepackage{booktabs}
\usepackage{lineno}
\usepackage{lipsum}
\usepackage{color}
\usepackage{hyperref}
\usepackage[nameinlink]{cleveref}
\usepackage{graphicx}% Include figure files
\usepackage{dcolumn}% Align table columns on decimal point
\usepackage{slashed}
\usepackage{comment}
\usepackage{wrapfig}
\usepackage{adjustbox}

\usepackage[font=small,labelfont=bf,justification=justified]{caption}
\usepackage[font=small,labelfont=bf, justification=centering]{subcaption}

\newcommand{\red}[1]{\textcolor{red}{#1}}

\title{The order of the chiral phase transition in massless many-flavour lattice QCD}
\author[a,b]{Reinhold Kaiser}
\author*[a]{Jan Philipp Klinger}
\author[a,b]{Owe Philipsen}

\affiliation[a]{Institute for Theoretical Physics - Goethe University, \\
Max-von-Laue-Str. 1, 60438 Frankfurt am Main, Germany}
\affiliation[b]{
    John von Neumann Institute for Computing (NIC) at GSI,
    \newline Planckstr. 1, 64291 Darmstadt, Germany
}

\emailAdd{klinger@itp.uni-frankfurt.de}
\emailAdd{kaiser@itp.uni-frankfurt.de}
\emailAdd{philipsen@itp.uni-frankfurt.de}

\abstract{The nature of the QCD phase transition in the chiral limit presents a challenging problem for lattice QCD. However, its study provides constraints on the phase diagram at the physical point. In this work, we investigate how the order of the chiral phase transition depends on the number of light quark flavours. To approach the lattice chiral limit, we map out and extrapolate the chiral critical surface that separates the first-order region from the crossover region in an extended parameter space, which includes the gauge coupling, the number of quark flavours, their masses, and the lattice spacing. Lattice simulations with standard staggered quarks reveal that for each $N_f < 8$, there exists a tricritical lattice spacing $a^\text{tric}(N_f)$, at which the chiral transition changes from first order ($a>a^\text{tric}$) to second order ($a<a^\text{tric}$). Thus, the first-order region is merely a lattice artifact and not connected to the continuum. By determining the associated temperatures  $T(N_f^\text{tric},a ^\text{tric})$ at these tricritical points, we confirm the expected decrease in the critical temperature as the number of flavours increases. The obtained temperatures define a tricritical line which is connected to the continuum and terminates at a physical $ N_f^\text{tric}(a=0) $. Our data is compatible with a vanishing temperature at that point, $T(N_f^\text{tric}(a=0))=0 $.}

\FullConference{The 41st International Symposium on Lattice Field Theory (LATTICE2024)\\
 28 July - 3 August 2024\\
Liverpool, UK\\}

\begin{document}
\maketitle

\section{Introduction}
\begin{figure}[t!]
   \begin{subfigure}{0.5\textwidth}
    \centering
        \includegraphics[width=0.9\linewidth]{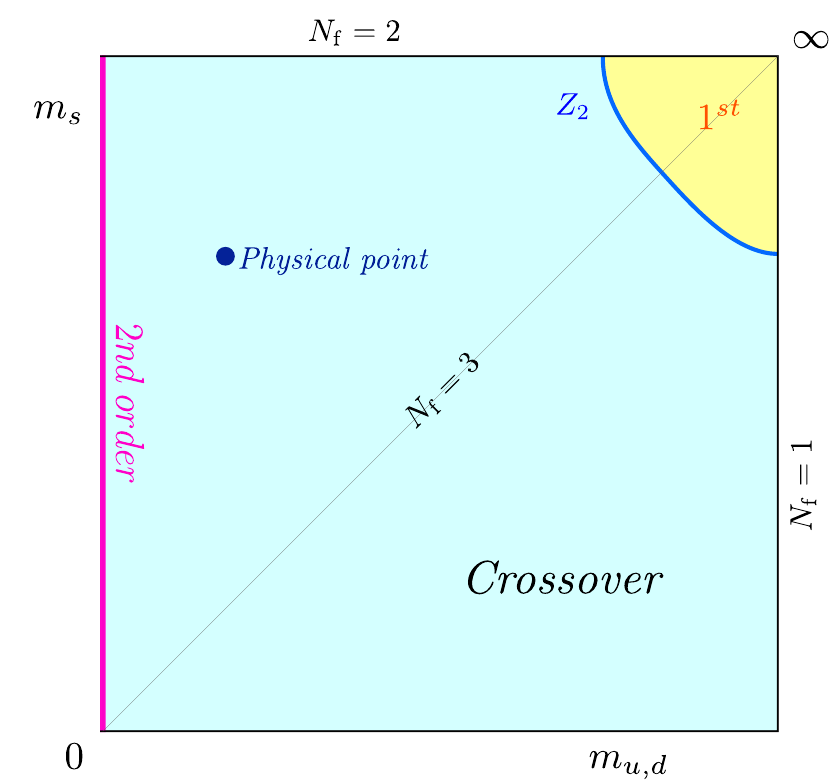}
    \caption{Columbia plot in plane of $m_s$ and $m_{u,d}$}
    \label{fig: Intro Illustration ms mud}  
        \end{subfigure}%
     \begin{subfigure}{0.5\textwidth}
    \centering
    \includegraphics[width=0.88\linewidth]{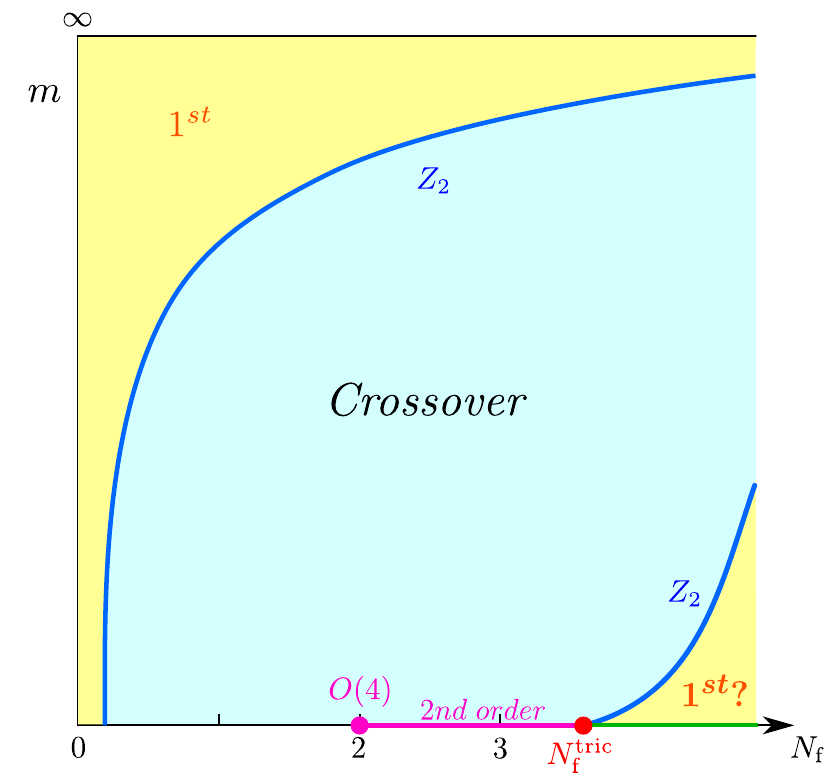}
    \caption{Columbia plot for mass-degenerate quarks}
    \label{fig: Intro Illustration m Nf}  
    \end{subfigure}
    \caption{Columbia plots. Every point represents a phase boundary with an implicitly associated (pseudo-) critical temperature $T_c$. Figures are taken from \cite{Cuteri_2021}.}
\end{figure}

\begin{figure}
\begin{minipage}[c]{0.49\linewidth}
\centering
\begin{subfigure}{1\textwidth}
\centering
\includegraphics[width=1\linewidth]{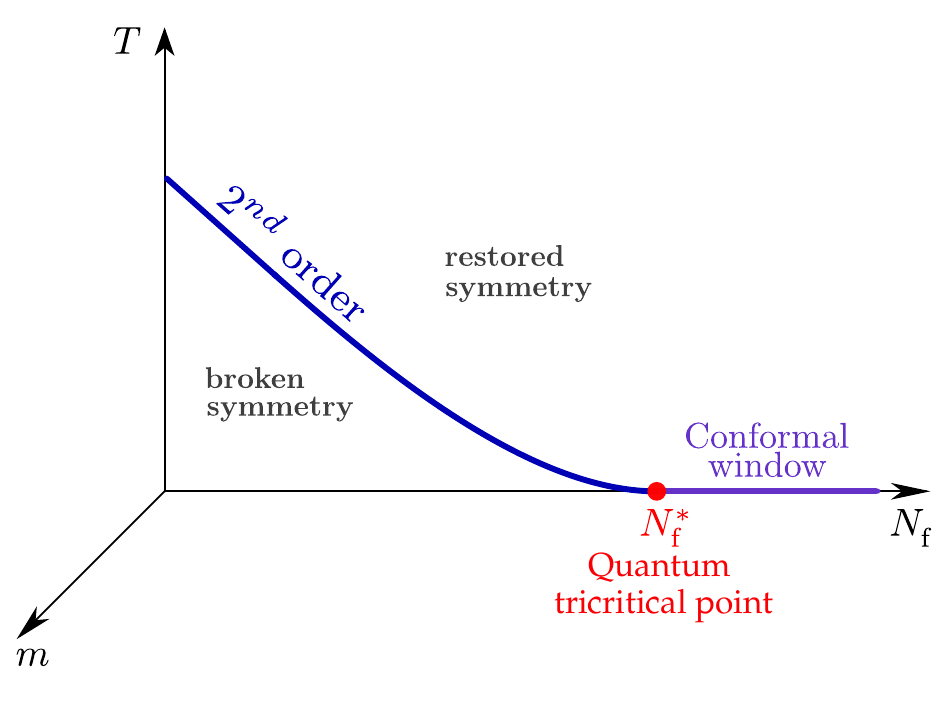}
    \fbox{$N_f^{\text{tric}}\;\;$ at $\;\;T=0$}
\caption{Szenario 1: 2nd-order transition for all $N_f$ }
    \label{fig: Intro Illustration T m Nf 2nd}  
\end{subfigure}
\end{minipage}\hfill
\begin{minipage}[c]{0.49\linewidth}
\centering
\begin{subfigure}{1\textwidth}
\centering
\includegraphics[width=1\linewidth]{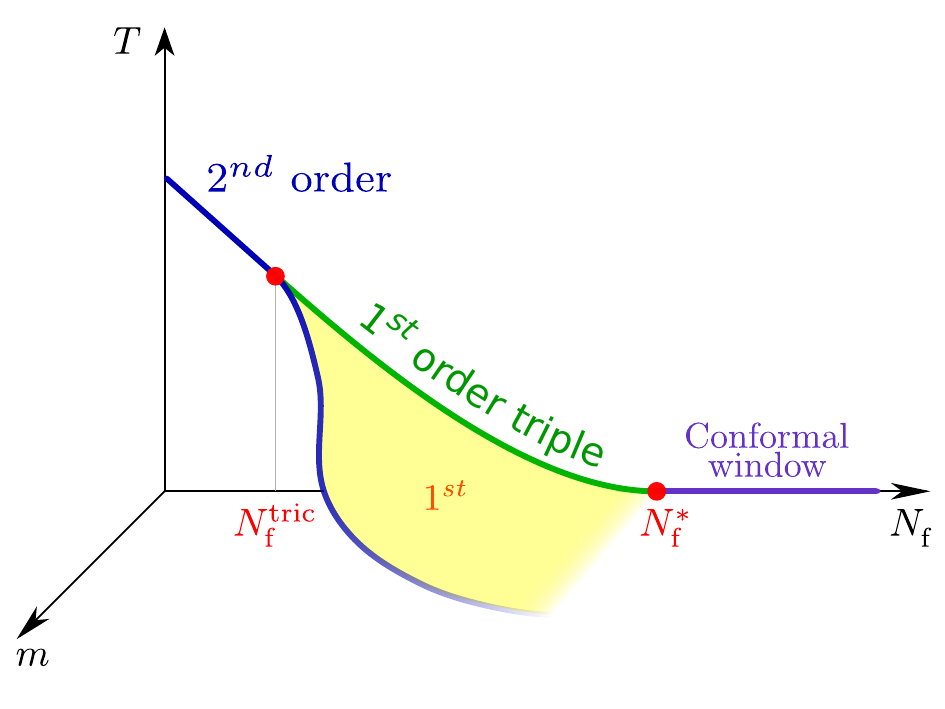}
    \fbox{$N_f^{\text{tric}}\;\;$ at $\;\;T>0$}
    \caption{Szenario 2: 1st-order transition for higher $N_f$}
        \label{fig: Intro Illustration T m Nf 1st}
\end{subfigure}
\end{minipage}
\caption{Comparison of phase diagrams for possible scenarios for the chiral limit depending on whether a first-order transition emerges for higher \( N_f \) or not. Figures are taken from \cite{Cuteri_2021}.}
        \label{fig: Intro Illustration T m Nf both}
\vspace{-3mm}
\end{figure}

The chiral limit refers to QCD in the presence of massless quarks. As a controllable deformation of QCD, it offers valuable insights into fundamental principles of the strong interaction and provides relevant constraints for physical QCD. Particularly with the aim of improving our understanding of the chiral phase transition, it is worth studying the massless limit of quarks. Only in the presence of massless quarks, the chiral symmetry is exact and thus its spontaneous breaking has to be accompanied by a true non-analytic phase transition.\\
How the chiral transition is affected by a change of the quark masses is illustrated in a so-called Columbia plot. Figure \ref{fig: Intro Illustration ms mud} depicts the nature of the QCD thermal transition as a function of degenerate up- and down-quark masses and the strange-quark mass \cite{Columbia}. The variation of masses can be used as an interpolation of QCD between one to three flavours. The chiral transition at the physical point is known to be an analytic crossover \cite{Aoki:2006we}. Quenched QCD in the limit of infinitely heavy masses (upper right corner) reduces to a $SU(3)$ Yang-Mills theory in the presence of static quarks and exhibits a first-order phase transition of the $\mathbb{Z}_3$-center symmetry \cite{BOYD1996419, Owe_Z3, Z2}. However, the situation in the chiral limit (lower and upper left corner) is more delicate. Massless quarks cannot be simulated directly with Monte-Carlo simulations due to zero modes in the dirac operator. Non-perturbative statements from first principles are therefore not straightforward. Nevertheless, an increased interest in recent years has led to accumulating confirmations that QCD with both, two and three flavours, exhibit a second-order transition for massless quarks. Support for this conclusion comes from lattice methods \cite{Cuteri_2021, Karsch, Dini:2021hug, wilson, Zhang:2022kzb, Zhang:2024ldl, Zhang:2025vns}, as well as functional approaches \cite{Fejos_2022, Fejos_2024, Braun_2023,Resch:2017vjs,Fischer_2023}.\\
The question remains whether a second-order transition persists in the chiral limit for high numbers of flavours. 
This is illustrated in Fig. \ref{fig: Intro Illustration m Nf}. Instead of using the strange mass as interpolation between two and three flavours, we consider degenerate quark masses and treat the number of fermions, $N_f$, as a continuous real parameter. Assuming now that a first-order transition emerges at some higher number of flavours, the chiral limit features triple points characterized by the coexistence of three distinct states (a vanishing, positive and negative chiral condensate at the critical temperature). The onset of the triple line is marked by a tricritical point. For a detailed description see \cite{Cuteri_2021}. All flavours below \( N_f^{\text{tric}} \) exhibit a second-order phase transition in the chiral limit, whereas those above undergo a first-order transition. The first-order region, which also extends to non-vanishing masses, is then bounded by a $\mathbb{Z}_2$-boundary line.
In fact, such a $\mathbb{Z}_2$-line was found by our group in lattice simulations \cite{Cuteri_2021, erstes_imag_mu}. However, it was simultaneously shown that the size of the first-order region decreases with decreasing lattice spacing. It was concluded that the first-order region is thus a cutoff effect, and the transition is second-order in the continuum limit for at least $N_f \leq 6$.\\
In this work we extend the Columbia plot (Fig. \ref{fig: Intro Illustration m Nf}) into a phase diagram, see Fig. \ref{fig: Intro Illustration T m Nf both}, by determining the critical temperatures. Of particular interest is the temperature at the tricritical point $N_f^\text{tric}$.
A rough outline of the expected behavior of the critical temperature as a function of the number of flavours in the chiral limit can be derived from perturbation theory. For small $N_f$ a linear decrease in \( T \) is predicted \cite{Braun:2009ns}, which transitions to exponential Miransky-scaling for higher \( N_f \) \cite{Miransky} and ultimately ends at the onset of the conformal window. The latter arises due to the emergence of non-trivial infrared (Banks-Zaks) fixed points \cite{BANKS}. The perturbative two-loop beta function of the running coupling suggests the fixed points to emerge between $N_f \simeq 8.05$ and $16.5$. However, non-perturbative dynamics might alter the onset $N_f^*$ of the conformal window, leaving it an open question for ongoing research. Studies suggest \( 8 \lesssim N_f^* \lesssim 12 \) \cite{Hasenfratz_10, Hasenfratz_12, Lombardo_12, Lombardo_8, Kotov:2021hri, Miura:2011mc, Lombardo, Braun:2006jd, Braun:2009ns}, with a growing tendency towards \( N_f^* = 8 \) being the sill of the conformal window \cite{Hasenfratz_8, Hasenfratz:2024zqn}. The absence of a running coupling renders QCD scale-invariant, resulting in a chirally symmetric phase and the lack of a thermal phase transition, i.e., \( T = 0 \).\\
Figure \ref{fig: Intro Illustration T m Nf both} illustrates this expected decrease of the critical temperature with the number of flavours. Fig. \ref{fig: Intro Illustration T m Nf 2nd} corresponds to the scenario that the second-order transition, found for $N_f=2$ and $3$, extends all the way down to $T = 0$, whereas Fig. \ref{fig: Intro Illustration T m Nf 1st} shows the potential opening of a first-order area, that is, the second-order line terminates at a finite T at $N_f^{\text{tric}}$ and is followed by a first-order transition up to the conformal window. Both scenarios have a tricritical point $N_f^\text{tric}$ but can be distinguished by its associated temperature at that point. While in scenario Fig. \ref{fig: Intro Illustration T m Nf 2nd}, the tricritical point coincides with the onset of the conformal window $N_f^*$ and has $T(N_f^{\text{tric}})=0$, the temperature at $N_f^{\text{tric}}$ in scenario Fig. \ref{fig: Intro Illustration T m Nf 1st} is $T(N_f^{\text{tric}}) > 0$.\\
Our goal is to determine \( N_f^{\text{tric}} \) and its corresponding temperature. Once $T(N_f^{\text{tric}})$ is known, it will reveal whether the conformal window is approached through a first-order or second-order transition. In case of the latter, our approach might even pinpoint the onset of the conformal window.

\nopagebreak
\section{Methodology and computational framework}

\paragraph{Strategy}\mbox{}\\
Since the chiral limit is not accessible to lattice simulations, it relies on extrapolations. If a first-order region is to emerge at higher \( N_f \) in the chiral limit, a tricritical point is guaranteed to exist which is approached by a second-order wingline with known scaling \cite{lawrie}
\begin{align}
\label{Eq: tricritical scaling in m}
N_f^c (m)= N_f^{\text{tric}} + A\cdot m^{2/5} + B\cdot m^{4/5} + \mathcal{O}(m^{6/5}).
\end{align}
Once the critical masses on this $\mathbb{Z}_2$-boundary line are determined for several $N_f$, the tricritical point $N_f^{\text{tric}}$ can be found by extrapolation. Nevertheless, as all simulations are performed on the lattice, the Columbia plots gets extended in another dimension consisting of the lattice spacing. In \cite{Cuteri_2021} it was found that the $\mathbb{Z}_2$-boundary highly depends on the lattice spacing.
We thus repeat mapping out the $\mathbb{Z}_2$-line and extracting $N_f^{\text{tric}}$ for several lattice spacings.

\paragraph{Simulation details}\mbox{}\\
Our QCD lattice simulations employ the standard Wilson gauge action and unimproved staggered fermions on lattices with size $N_\tau \times N_\sigma^3$. The tuneable bare parameters are the degenerate quark mass $am$, the inverse gauge coupling  $\beta=6/g^2$ and the number of fermions $N_f$. The coupling $\beta$ controls the lattice spacing $a$ and tunes the temperature through the relation $T=1/[a(\beta)N_\tau]$. By keeping $T$ constant, the lattice spacing can be reduced by increasing $N_\tau$. Our codebase is built on the OpenCL-based lattice QCD framework CL2QCD \cite{cl2qcd} . It is executed on the GPU clusters VIRGO at GSI in Darmstadt and Goethe-HLR at the Center for Scientific Computing in Frankfurt.

\paragraph{Determining the critical surface}\mbox{}\\
A critical point on the $\mathbb{Z}_2$-boundary corresponds to a set of critical couplings $\{\beta^{\mathbb{Z}_2}_c, am^{\mathbb{Z}_2}_c\}$ at fixed $N_f$ and $N_\tau$. The order of the chiral transitions is studied by the use of the chiral condensate as a (quasi-)order parameter, $\langle \mathcal{O} \rangle =  \langle \bar{\Psi} \Psi \rangle$, and its distribution is analysed via its generalised moments
\begin{align}
    B_n = \frac{ \left\langle \left( \mathcal{O} - \langle \mathcal{O} \rangle \right)^n \right\rangle\;\;\;\;\;}{\left\langle \left( \mathcal{O} - \langle \mathcal{O} \rangle \right)^2 \right\rangle^{n/2}}.
\end{align}
To obtain the critical coupling $\beta^{\mathbb{Z}_2}_c$ and critical mass $am_c^{\mathbb{Z}_2}$ at some fixed $N_f$ and $N_\tau$, we perform a finite size scaling analysis. We start to determine the (pseudo-)critical $\beta_{pc}$ at three different masses in the vicinity of the critical mass value $am^{\mathbb{Z}_2}_c$ by scanning in the lattice gauge coupling for vanishing skewness, $B_3(\beta_{pc}, am, N_\sigma) = 0$. This is repeated for three different aspect ratios $N_\sigma/N_\tau \in \{2,3,4\}$.
Identifying the kurtosis $B_4(\beta_{pc}, am, N_\sigma)$ on this pseudocritical hypersurface and assuming its associated critical value of $B_4^{\mathbb{Z}_2} = 1.6044(10)$, the $\mathbb{Z}_2$-critical mass is extracted by a finite size scaling fit 
\begin{align}
\label{Finite size scaling}
B_4(\beta_{pc}, am, N_\sigma) \approx \left(B_4^{\mathbb{Z}_2} + c[am-am^{\mathbb{Z}_2}_c]N_\sigma^{1/v} \right)\left( 1+ b N_\sigma^{y_t-y_h} \right).
\end{align}
The last factor is a finite volume correction term, where $y_t = 1 /v = 1.5870(10)$ and $y_h  = 2.4818(3)$ are the associated 3D Ising exponents \cite{Ising, scaling1}. Once the critical mass $am^{\mathbb{Z}_2}_c$ is known, the critical values of the coupling constant $\beta^{\mathbb{Z}_2}_c$ at $am^{\mathbb{Z}_2}_c$ are obtained by a linear fit of the (pseudo-)critical $\beta_{pc}$ values at the simulated quark masses.

\section{Results}

\subsection{Phase boundary in the lattice parameter space}

\begin{figure}[t!]
   \begin{subfigure}[t]{0.5\textwidth}
    \centering
        \raisebox{+0.04cm}{\includegraphics[height=4.78cm]{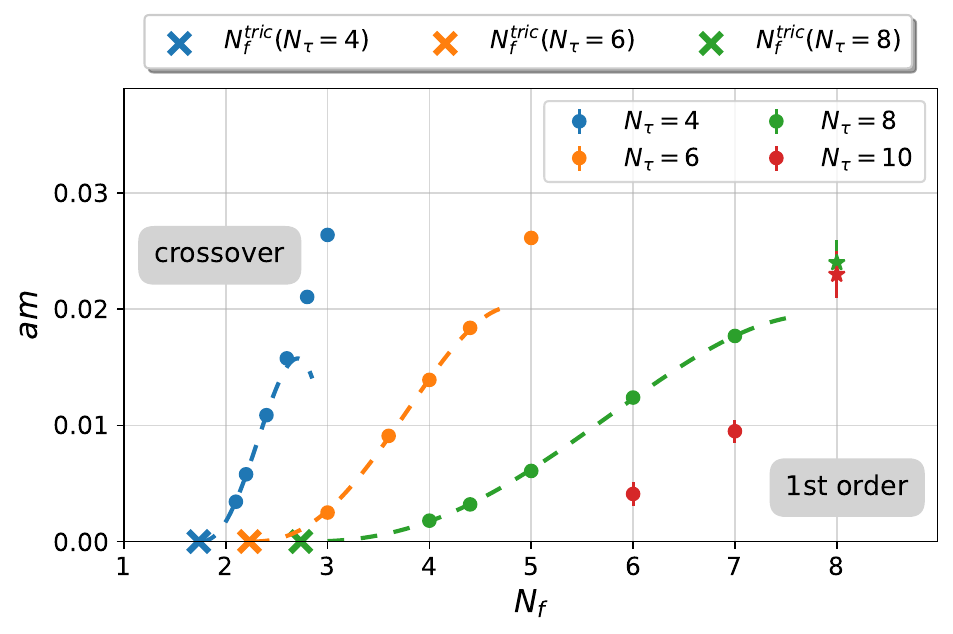}}
    \caption{Projections with fixed $N_\tau$}
    \label{fig: Results am Nf}  
        \end{subfigure}%
     \begin{subfigure}[t]{0.5\textwidth}
    \centering
   {\includegraphics[height=4.8cm]{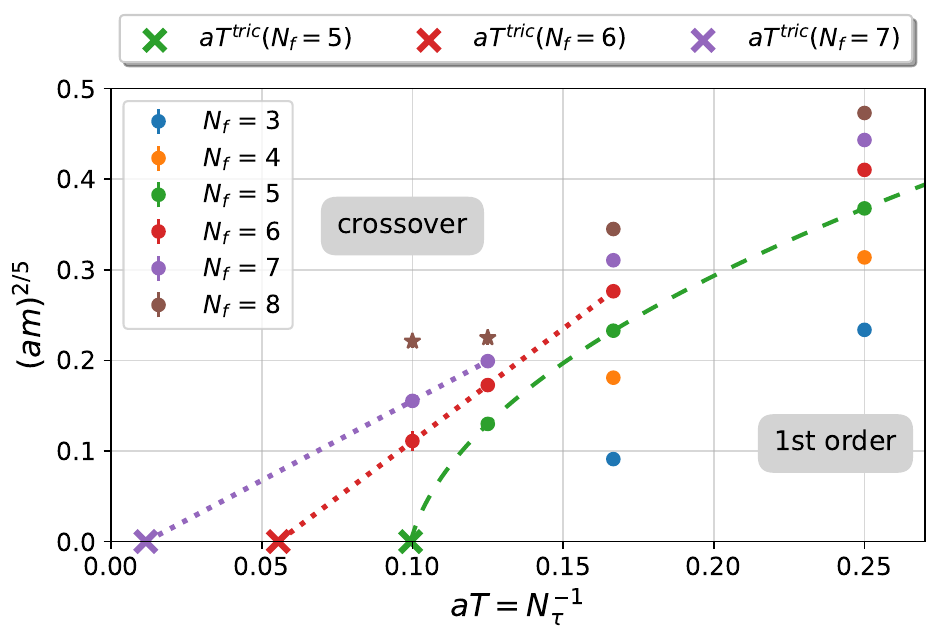}}      
    \caption{Projections with fixed $N_f$}
    \label{fig: Results am aT}  
    \end{subfigure}
    \caption{The $\mathbb{Z}_2$-boundary $\{\beta_c^{\mathbb{Z}_2},am_c^{\mathbb{Z}_2}, N_f, N_\tau \}$ projected onto different planes. Every point represents a phase boundary with an implicitly tuned $\beta_c^{\mathbb{Z}_2}(am, N_f, N_\tau)$. The lines correspond to fits according to Eq.(\ref{Eq: tricritical scaling in m}) for the left panel and Eq.(\ref{Eq: tricritical scaling for aT}) for the right panel. On the right, dotted lines are LO-fits, while dashed lines are NLO-fits. The points for $N_f=8$ at $N_\tau=8$ and $10$ do not correspond to the $\mathbb{Z}_2$-boundary, but presumably mark the onset of the lattice bulk transition, see section \ref{Sec: Bulk Transition}. } 
\end{figure}

The chiral critical surface that separates the first-order region from the crossover has been mapped out in an enlarged parameter space of our lattice action $\{\beta, am, N_f, N_\tau\}$. For several numbers of flavours $N_f \in [2,8]$ and lattice spacings $N_\tau \in \{4,6,8,10\}$ the critical couplings $\{\beta_c^{\mathbb{Z}_2}, am_c^{\mathbb{Z}_2}\}$ were identified, characterizing the $\mathbb{Z}_2$-boundary line. Figure \ref{fig: Results am Nf} shows the critical masses $am^{\mathbb{Z}_2}_c$ over the number of flavours; the values of the critical coupling are implicit. This figure is the lattice version of Fig. \ref{fig: Intro Illustration m Nf} and illustrates how the first-order region behaves with the lattice spacing $N_\tau$. The indicated lines correspond to the $\mathbb{Z}_2$-boundary fitted according to Eq.(\ref{Eq: tricritical scaling in m}). It can be seen that the first-order region -- masses below the respective lines -- is highly cutoff dependent and shrinks for decreasing lattice spacing. Furthermore, the onset of the first-order region $N_f^{\text{tric}}(N_\tau)$ is pushed to higher $N_f$, the lower the lattice spacing.
This analysis eventually raises the question, whether a first-order region even remains for any number of flavours in the continuum. This is tested in Fig. \ref{fig: Results am aT}, where $am$ is plotted over $N_\tau^{-1}=aT$. Note, that the continuum limit corresponds to the origin of the plot as $am \rightarrow 0$ and $N_\tau \rightarrow \infty$. For the first-order region to be physical, the $\mathbb{Z}_2$-boundary line must hence connect to the origin of the plot, i.e., the continuum. In this variable pair tricritical scaling of the $\mathbb{Z}_2$-boundary takes the form  
\begin{align}
\label{Eq: tricritical scaling for aT}
aT^c (am, N_f)= aT^\text{tric}(N_f) + A(N_f) (am)^{2/5} +  B(N_f) (am)^{4/5}+\mathcal{O}\left( (am)^{6/5} \right) ,
\end{align}
for each $N_f$. In Fig. \ref{fig: Results am aT} we restricted the fit range to smaller masses, as we expect stronger scaling behaviour. For $N_f = 3$ and $4$, we only have two data points making an extrapolation impossible. A clear support of tricritical scaling is given by \(N_f = 6\) as it exhibits leading-order scaling across three lattice spacings. The case of \(N_f = 8\) represents an exception and will be discussed in detail in Section \ref{Sec: Bulk Transition}. For each number of flavours, the intersection with the x-axis yields the tricritical point \(aT^{\text{tric}}(N_f)\). This point marks the tricritical lattice spacing in the chiral limit where the chiral transition changes from first order ($aT>aT^\text{tric}$) to second order ($aT<aT^\text{tric}$). As our extrapolations terminate for all flavours \(N_f \leq 7\) at non-zero $aT^\text{tric}(N_f)$ -- and $not$ in the origin -- we conclude that for all flavours \(N_f \leq 7\), the first-order region is merely a cutoff effect, and the transition in the continuum chiral limit is second order for all \(N_f \leq 7\). \\
Accordingly, the lattice theory differs significantly from continuum QCD in qualitative terms. We found that on the lattice, every number of flavours, \(N_f \leq 7\), exhibits a tricritical point at some \(aT^{\text{tric}}(N_f)\), with a clear trend of \(aT^{\text{tric}} \rightarrow 0\) as \(N_f\) increases. In contrast, continuum QCD features only one unique tricritical point, \(N_f^{\text{tric, phys}}\) at $a=0$. Based on the qualitative behaviour shown in Fig. \ref{fig: Results am aT}, this $physical$ \(N_{f}^{\text{tric, phys}}\) corresponds to a (not necessarily integer) value, where the tricritical line terminates at the origin of the plot, that is, \(aT^{\text{tric}}(N_f^{\text{tric, phys}}) = 0\).
Our data suggest that this physical tricritical point must be located at $N_f^{\text{tric, phys}}>7$.
Whether this point signals the onset of first-order transitions or marks the end of chiral symmetry breaking, i.e., the beginning of the conformal window, is revealed by the corresponding temperature at that point.

\subsection{Phase boundary in physical units}
\begin{figure}
\begin{minipage}[t]{0.47\linewidth}
\centering
{\includegraphics[height=4.8cm]{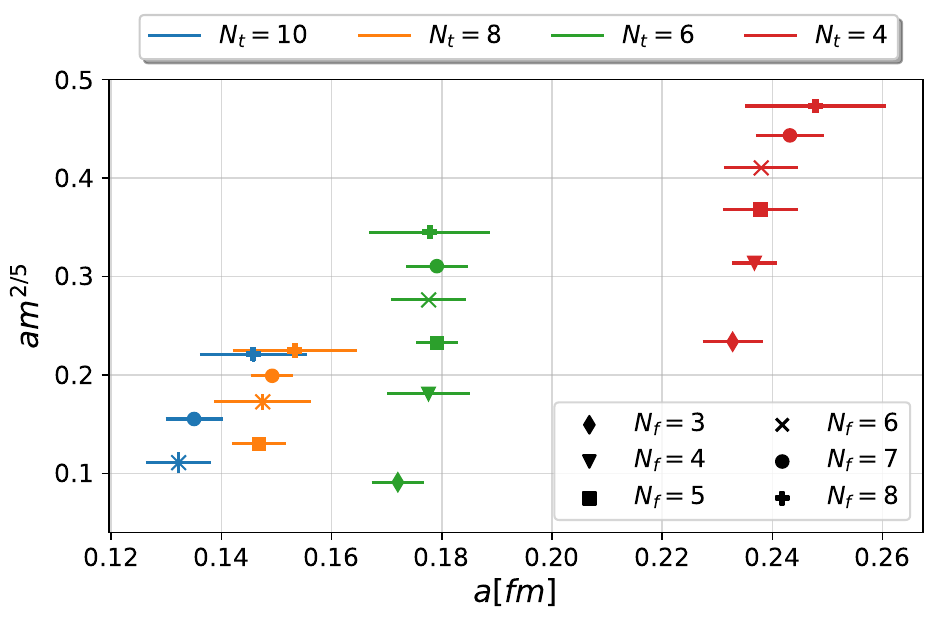}}
\caption{\label{fig: Results am a}The lattice spacings of the $\mathbb{Z}_2$-boundary in physical units.}
\end{minipage}%
\hfill
\begin{minipage}[t]{0.47\linewidth}
    \centering
\raisebox{0.02cm}{\includegraphics[height=4.8cm]{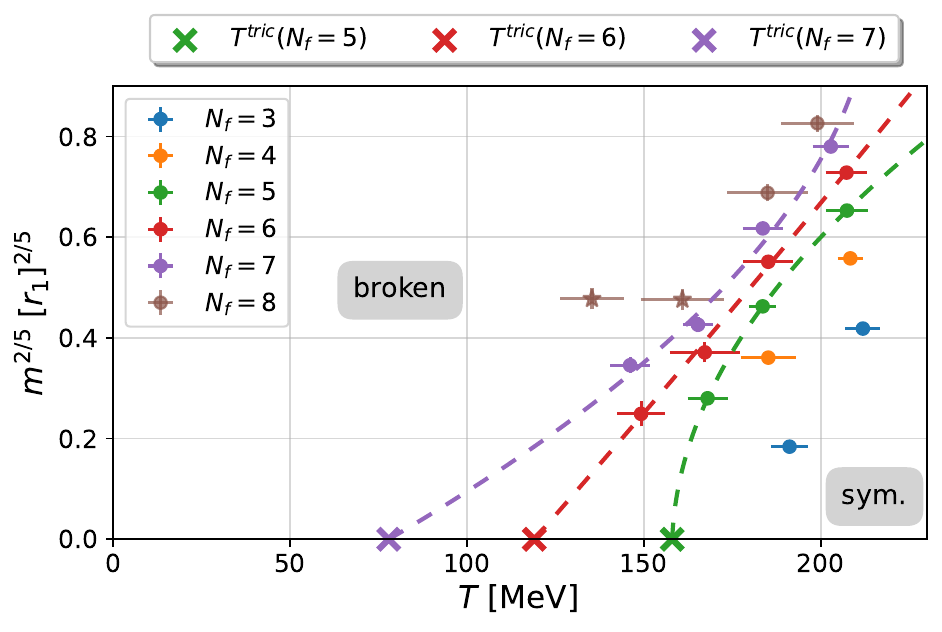}}
    \caption{\label{fig: Results m T}The critical temperatures on the $\mathbb{Z}_2$-boundary. The lines correspond to fits according to Eq.(\ref{Eq: tricritical scaling for T}). The points for $N_f=8$ at $N_\tau=8$ and $10$ do not correspond to the $\mathbb{Z}_2$-boundary.}
\end{minipage}
\end{figure}

\begin{figure}
    \vspace{-4mm}
\centering
    \includegraphics[width=0.5\linewidth]{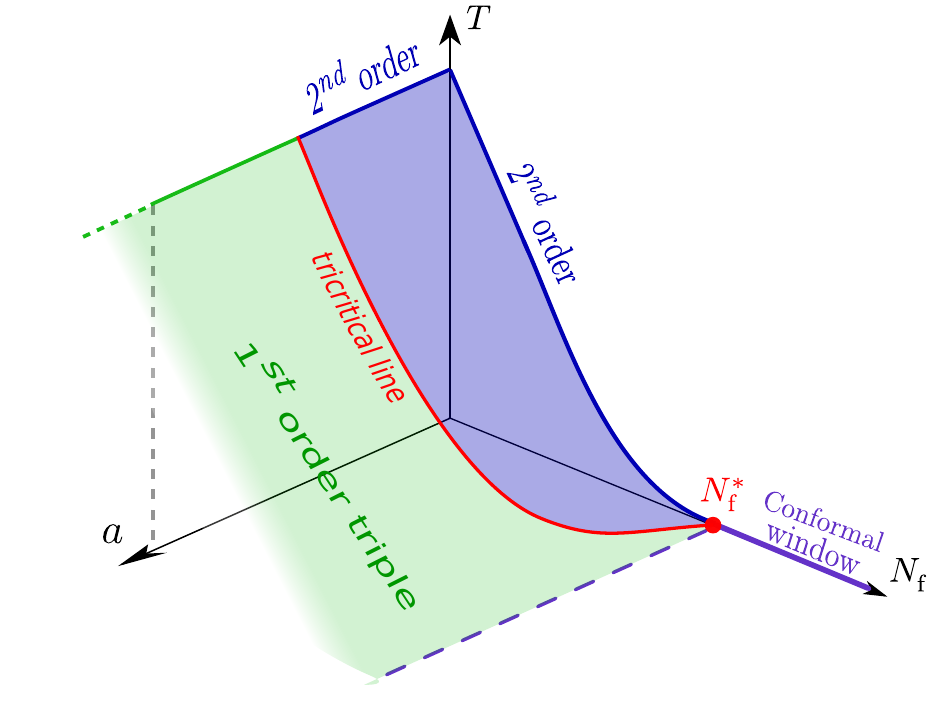}
    \vspace{-2mm}
    \caption{Lattice phase diagram in the chiral limit. Tricritical points $\{N_f^\text{tric}, T^\text{tric}, a^\text{tric} \}$ in red separate the first order from the second-order region. Only the latter is connected to the continuum $a=0$.   }
    \label{fig: Ill T a Nf} 
    \vspace{-2mm}
\end{figure}

The temperature is calculated by introducing physical units through setting the scale in our lattice simulations. We stress that comparing QCD with different numbers of flavours requires caution as this is not merely a modification of standard bare parameters but rather a change of the entire theory. A fixed reference scale for different $N_f$ is hence problematic.  Furthermore, the understanding of units loses its conventional meaning in this setting, as our study with degenerate quarks and various $N_f$ values departs significantly from the physical point.
Our method of choice for measuring the lattice spacing is via the improved Sommer parameter $r_1$ \cite{Sommer_r0, Sommer_new}. It is directly tied to the force between two static quarks and has proven to be robust against changes in quark mass and number of flavours \cite{Bruno_sommer, Knechtli}.\\
The calculated lattice spacings for our data of the $\mathbb{Z}_2$-boundaries are shown in Fig. \ref{fig: Results am a} in units of $fm$. We chose this representation similar to Fig. \ref{fig: Results am aT}, with $am$ over $aT = N_\tau^{-1}$, as the connection $a \sim N_\tau^{-1}$ becomes directly evident. Once $a$ is known, all other lattice observables can be converted to physical units. Thus, we derive Fig. \ref{fig: Results m T} showing the mass in units of $r_1$ over the critical temperature in $MeV$. Again, we perform fits for each $N_f$ according to our tricritical scaling ansatz
\begin{align}
\label{Eq: tricritical scaling for T}
T^c (m, N_f)= T^\text{tric}(N_f) + A(N_f) m^{2/5} + B(N_f)m^{4/5}+ \mathcal{O}(m^{6/5}).
\end{align}
The tricritical points $T^\text{tric}(N_f)$, i.e., the intersections with the x-axis, tell us at which temperature chiral symmetry is restored in the chiral limit. Note that these are not the critical temperatures in the continuum, but rather the transition temperatures at the tricritical points which, demonstrated in the previous section, are at finite lattice spacings $aT^\text{tric}$. Nevertheless, Fig. \ref{fig: Results m T} reveals a decreasing trend in the tricritical temperature, with \(T^{\text{tric}}(N_f) \rightarrow 0\) as \(N_f\) increases. 
In combination with the finding of the previous section, namely $aT^\text{tric}(N_f) \rightarrow 0$, our data is consistent with the existence of a unique (not necessarily integer) tricritical \(N_f^\text{tric, phys}\) in the continuum with a temperature of $T(N_f^\text{tric, phys})=0$. The thus resulting picture is visualized in Fig. \ref{fig: Ill T a Nf} and discussed in the conclusion.

\subsection{Bulk Transition for $\boldsymbol{N_f = 8}$}
\label{Sec: Bulk Transition}
The analysis based on tricritical scaling for numbers of flavours $N_f\leq 7$ cannot be applied for $N_f=8$. The reason is the occurrence of the lattice bulk transition at $N_\tau =8$ and $10$. This transition is a pure lattice artifact and does not correspond to a physical phase transition in the continuum. It is associated with the discretization of spacetime, not the physical continuum system, and becomes more pronounced at strong coupling (large lattice spacing). See \cite{Deuzeman:2010xqg, Deuzeman:2011pa, Deuzeman:2012ee, hasenfratz_bulk} for details and references. The bulk transition differs from a thermal phase transition in that it is independent of the temperature, or $N_\tau$.  This is observed in Figures \ref{fig: Results am Nf} and \ref{fig: Results am aT}, where increasing $N_\tau$ from $8$ to $10$ does not result in a change in the critical mass for $N_f=8$. Furthermore, the critical $\beta$-values for $N_\tau = 8$ and $10$ are also identical (not shown). This is eventually the reason for the similar lattice spacings  in Fig. \ref{fig: Results am a},  even though simulations were performed at different values of $N_\tau$. It suggests that the critical masses at $N_\tau = 8$ and $10$ do not correspond to a $\mathbb{Z}_2$-boundary, but to the onset of the bulk transition. Determining the critical mass-value $m^{\mathbb{Z}_2}_c$ of the thermal transition is no longer within reach, as for $N_\tau \geq 8$, it falls below the onset of the bulk transition. For masses below the onset, we are in a bulk regime where the thermal restoration of chiral symmetry ceases to exist. Concluding, for $N_f=8$, the thermal $\mathbb{Z}_2$-line does not terminate in the chiral limit at a tricritical point, but instead ends at a non-zero mass $am^\text{bulk}>0$ which is the onset of the bulk phase. There is hence no basis for tricritical scaling. Nevertheless, based on the fact that it is well established that the bulk regime is a lattice artifact and not physical, we conclude that for \(N_f = 8\), the observed first-order region is again not connected to the continuum.

\begin{comment}
However, one can draw the same conclusion for $N_f=8$ as before. 
We argued that, for each $N_f \leq 7$, the first-order region only exists for large lattice spacings and ends at a certain \(a^{\text{tric}}(N_f)>0\). For \(N_f \geq 8\), we find that the thermal first-order region merges with the one of the bulk transition. Based on the fact that it is well established that the latter is a lattice artifact and not physical, we conclude that for \(N_f = 8\), the first-order region is also not connected to the continuum. \red{todo: references}
\end{comment}

\section{Conclusion}
\vspace{-2mm}
For unimproved staggered fermions, a first-order transition exists in the chiral limit for all \( N_f \) on coarse lattices. However, this is merely a cutoff effect and terminates at a finite tricritical lattice spacing, \( N_\tau^{\text{tric}^{-1}} =aT^\text{tric}(N_f) \), for all \( N_f \leq 7 \). It was observed that \( aT^\text{tric} \rightarrow 0 \) as \( N_f \) increases. The temperatures at these tricritical points were determined through scale setting, revealing that \( T^\text{tric}(N_f) \rightarrow 0 \) with increasing \( N_f \). The resulting picture is presented in Fig. \ref{fig: Ill T a Nf}. The $T-N_f$ phase diagram in the chiral limit is extended by a third dimension being the lattice spacing. 
The indicated surface separates a chirally broken regime (below) from a chirally symmetric regime (above). For each lattice spacing $a$ there is a corresponding tricritical point $N_f^\text{tric}(a)$. The tricritical points lie on a tricritical line indicated in red. This line marks the separation between the first-order region for large lattice spacing and a second-order region for small lattice spacing. Following the tricritical line to $a=0$, it approaches the physical tricritical point $N_f^\text{tric, phys}$. Our data is consistent with a physical tricritical point at $T=0$. In this case $N_f^\text{tric, phys}$ coincides with the onset of the conformal window $N_f^*$ and marks the end of chiral symmetry breaking. Accordingly, a first-order transition in the chiral limit can be excluded for all numbers of flavours. We have not yet been able to pinpoint \( N_f^\text{tric, phys} \). However, we have first indications for a non-integer value between \( 7 < N_f^\text{tric, phys} < 8 \). This physical \(N_f^\text{tric, phys}\) is then characterized by a $\mathbb{Z}_2$-boundary line terminating, on the one hand, in the origin of Fig. \ref{fig: Results am aT}, i.e., $aT^\text{tric}=0$ and, on the other hand, also in the origin of Fig. \ref{fig: Results m T}, i.e., $T^\text{tric}=0$. For \( N_f = 8 \), we observe a lattice bulk transition and the absence of a tricritical point. As the bulk transition is known to be a discretization artefact, the statement remains valid that the first-order region observed on the lattice is not connected to the continuum for any \( N_f \).  This renders the transition in the chiral limit second order for all numbers of flavours unless a, so far unknown, first-order region emerges at very small lattice spacings.
\begin{comment}
\begin{wrapfigure}{r}{0.5\textwidth}
\centering
    \includegraphics[width=1\linewidth]{Figures/3_Results/T_a_nf_2st.pdf}
    \caption{Lattice phase diagram in the chiral limit. Tricritical points $\{N_f^\text{tric}, T^\text{tric}, a^\text{tric} \}$ in red separate the first order from the second-order region. Only the latter is connected to the continuum $a=0$.  }
    \label{fig: Ill T a Nf}  
\end{wrapfigure}
\end{comment}

\begin{comment}
In our previous argument, we assumed that the onset of the bulk transition is greater than or equal to the onset of the conformal window. If it were to turn out that the bulk transition interferes with the tricritical scaling at even lower $N_f$ (possibly at higher \( N_t \)), this would prevent the determination of \( N_f^\text{tric, phys} \) and, consequently, the onset of the conformal window.
\end{comment}

\acknowledgments{ \vspace{-2mm} This work is supported by the Deutsche Forschungsgemeinschaft (DFG) through the grant
CRC-TR 211 and by the State of Hesse within the Research Cluster ELEMENTS. We thank the
staff of VIRGO at GSI Darmstadt for computing resources and acknowledge use of the analysis
software packages “Monte Carlo Cpp analysis tools” by A. Sciarra et al. and “PLASMA” by C. Pinke et al..}

\bibliographystyle{JHEP}
\bibliography{bibliography}

%\begin{thebibliography}{99}
%\bibitem{...}
%\end{thebibliography}

\end{document}